\documentclass[showpacs,twocolumn]{revtex4}
\usepackage{amsmath}
\usepackage{amssymb}
\usepackage{graphicx}
\usepackage{bm}
\usepackage{dcolumn}

\begin{document}

\title{Monotonic growth of interlayer magnetoresistance in strong magnetic
field in very anisotropic layered metals}
\author{P. D. Grigoriev}
\email{grigorev@itp.ac.ru}
\affiliation{L. D. Landau Institute for Theoretical Physics, Chernogolovka, Russia}
\date{\today }

\begin{abstract}
It is shown, that the monotonic part of interlayer electronic conductivity
strongly decreases in high magnetic field perpendicular to the conducting
layers. We consider only the coherent interlayer tunnelling, and the
obtained result strongly contradicts the standard theory. This effect
appears in very anisotropic layered quasi-two-dimensional metals, when the
interlayer transfer integral is less than the Landau level separation.
\end{abstract}

\maketitle

\textbf{Introduction.} The investigation of the angular and magnetic field
dependence of magnetoresistance provides a powerful tool of studying the
electronic properties of various metals. The Fermi surface geometry of the
most metals has been measured using the magnetic quantum oscillations (MQO)
of magnetoresistance.\cite{Shoenberg,Abrik,Ziman} The angular dependence of
magnetoresistance also gives the important information about the electronic
structure and is widely used to investigate the electronic properties of
layered compounds: organic metals (see, e.g., Refs. \cite%
{KartPeschReview,MarkReview2004,OMRev,MQORev} for reviews), cuprate
high-temperature superconductors,\cite%
{HusseyNature2003,AbdelNature2006,AbdelPRL2007AMRO,McKenzie2007}
heterostructures\cite{Kuraguchi2003} etc.

In layered quasi-2D metals, where the interlayer transfer integral $t_{z}$
is considerably smaller than the in-plane electron Fermi energy, the
electron dispersion is given in the tight-binding approximation by
\begin{equation}
\epsilon _{3D}\left( \mathbf{k}\right) \approx \epsilon \left(
k_{x},k_{y}\right) -2t_{z}\cos (k_{z}d),  \label{ES3D}
\end{equation}%
where $\epsilon \left( k_{x},k_{y}\right) $ is the in-plane electron
dispersion, $k_{z}$ is out-of-plane electron momentum, and $d$ is the
interlayer spacing. If $t_{z}$ still much larger than the Landau level (LL)
separation $\hbar \omega _{c}=\hbar eB/m^{\ast }c$, the standard theory of
galvanomagnetic properties\cite{Abrik,Shoenberg,Ziman} works well. This
theory predicts several special features of magnetoresistance in quasi-2D
metals: the angular magnetoresistance oscillations\cite{Yam,Yagi1990} and
the beats of the amplitude of MQO.\cite{Shoenberg}

In strongly anisotropic layered quasi-2D metals, when $t_{z}\sim \hbar
\omega _{c}$, many new qualitative effects emerge. For example, the slow
oscillations of magnetoresistance appear\cite{SO,Shub} and the beats of MQO
of transport quantities become shifted.\cite{PhSh,Shub} These effects are
not described by the standard theory,\cite{Shoenberg,Abrik,Ziman} because it
is valid only in the lowest order in the parameter $\hbar \omega _{c}/t_{z}$%
. When this parameter becomes of the order of unity, the standard theory is
no longer applicable.

The monotonic part of magnetoresistance also changes when $t_{z}\lesssim
\hbar \omega _{c}$. According to the standard theory,\cite{Abrik} external
magnetic field along the electric current leads only to MQO but does not
influence the monotonic (background) part of this current. However, the
monotonic increase of interlayer magnetoresistance $R_{zz}$ with the
increase the magnetic field $\mathbf{B}$ perpendicular to the conducting
layers has been observed in various strongly anisotropic layered metals.\cite%
{Zuo1999,Kang,W3,W4,Wang2005,Incoh2009,Wosnitza2002} This monotonic growth
of magnetoresistance was attributed to the "strongly incoherent" regime,
where the interlayer tunnelling described by the usual Hamiltonian term in
Eq. (\ref{Ht}) is not effective, and the new mechanisms of interlayer
electron transport play the major role. For example, the variable-range
electron hopping between the localized states in strong magnetic field leads
to the insulating behavior and to the exponential dependence of interlayer
conductivity on temperature and magnetic field.\cite{Gvozd2007} In another
model, where the in-plane electron motion is fully metallic but the
interlayer electron transport goes via rare local crystal defects (e.g.,
resonance impurities), the interlayer conductivity $\sigma _{zz}$ also has
metallic-type temperature dependence but decreases strongly with the
increase of the out-of-plane component of magnetic field.\cite{Incoh2009}
The boson-assisted interlayer tunnelling can describe only the unusual
temperature dependence of interlayer conductivity at $T\sim 10-150K$,\cite%
{Lundin2003,Ho,Maslov} but it does not explain its magnetic field
dependence. Below I show, that the monotonic growth of magnetoresistance $%
R_{zz}\propto \sqrt{B_{z}}$ appears also in the standard model, described by
the Hamiltonian in Eqs. (\ref{H})-(\ref{Hi}), in strong magnetic field at
very weak interlayer coupling: $\hbar \omega _{c}\gg \Gamma _{0}>t_{z}$,
where $\Gamma _{0}=\hbar /2\tau _{0}$ is the electron level broadening due
to impurity scattering in the absence of magnetic field and $\tau _{0}$ is
the electron mean free time. This contradicts the common opinion\cite%
{MosesMcKenzie1999} that in the "weakly incoherent" regime, i.e. at $\Gamma
_{0}>t_{z}$, the interlayer magnetoresistance does not differ from the
coherent almost 3D limit $t_{z}\gg \Gamma _{0}$. This increase of
magnetoresistance was also missed in Refs. \cite%
{ChampelMineev,Gvozd2004,ChMineevComment2006}, where the Born approximation
has been incorrectly applied to describe almost the 2D electron system.

\medskip

\textbf{The model.} The electron Hamiltonian in layered compounds with small
interlayer coupling contains 3 main terms:
\begin{equation}
\hat{H}=\hat{H}_{0}+\hat{H}_{t}+\hat{H}_{I}.  \label{H}
\end{equation}%
The first term $\hat{H}_{0}$ is the noninteracting 2D electron Hamiltonian
summed over all layers:
\begin{equation}
\hat{H}_{0}=\sum_{m,j}\varepsilon _{2D}\left( m\right) c_{m,j}^{+}c_{m,j},
\label{H0}
\end{equation}%
where $\left\{ m\right\} =\left\{ n,k_{y}\right\} $\ is the set of quantum
numbers of electrons in magnetic field on a 2D conducting layer, $%
c_{m,j}^{+}(c_{m,j})$ are the electron creation (annihilation) operators in
the state $\left\{ m\right\} $ on the layer $j$, and $\varepsilon
_{2D}\left( m\right) $\ is the corresponding free electron dispersion given
by%
\begin{equation}
\varepsilon _{2D}\left( n,k_{y}\right) =\hbar \omega _{c}\left( n+1/2\right)
.  \label{En2D}
\end{equation}%
The second term in Eq. (\ref{H}) gives the coherent electron tunnelling
between two adjacent layers:
\begin{equation}
\hat{H}_{t}=2t_{z}\sum_{j}\int d^{2}\boldsymbol{r}[\Psi _{j}^{\dagger }(%
\boldsymbol{r})\Psi _{j-1}(\boldsymbol{r})+\Psi _{j-1}^{\dagger }(%
\boldsymbol{r})\Psi _{j}(\boldsymbol{r})],  \label{Ht}
\end{equation}%
where $\Psi _{j}(\boldsymbol{r})$ and $\Psi _{j}^{\dagger }(\boldsymbol{r})$%
\ are the creation (annihilation) operators of an electron on the layer $j$
at the point $\boldsymbol{r}$. This interlayer tunnelling Hamiltonian is
called "coherent" because it conserves the in-layer coordinate dependence of
the electron wave function (in other words, it conserves the in-plane
electron momentum) after the interlayer tunnelling. The last term\
\begin{equation}
\hat{H}_{I}=\sum_{i}\int d^{3}\mathbf{r}V_{i}\left( \mathbf{r}\right) \Psi
^{\dagger }(\boldsymbol{r})\Psi (\boldsymbol{r})  \label{Hi}
\end{equation}%
gives the electron interaction with impurity potential. The impurities are
taken to be point-like and randomly distributed on conducting layers with
volume concentration $n_{i}$ and areal concentration $N_{i}=n_{i}d$ on each
layer. The impurity distributions on any two adjacent layers are
uncorrelated. The potential $V_{i}\left( \mathbf{r}\right) $ of any impurity
located at point $\mathbf{r}_{i}$ is given by
\begin{equation}
V_{i}\left( \mathbf{r}\right) =U\delta ^{3}\left( \mathbf{r}-\mathbf{r}%
_{i}\right) .  \label{Vi}
\end{equation}%
We also introduce the 2D point-like impurity potential with the strength $%
V_{0}=U\left\vert \psi \left( z_{i}\right) \right\vert ^{2}\approx U/d$ of
each impurity:
\begin{equation}
V_{i}\left( x,y\right) =V_{0}\delta \left( x-x_{i}\right) \delta \left(
y-y_{i}\right) .  \label{Vi2D}
\end{equation}

In the limit, $t_{z}\ll \Gamma _{0},\hbar \omega _{c}$, the interlayer
hopping $t_{z}$ must be considered as a perturbation for the disordered
uncoupled stack of 2D metallic layers. The 2D metallic electron system in
magnetic field in the point-like impurity potential has been extensively
studied.\cite{Ando,Ando1,Baskin,Brezin,QHE,Imp,Burmi} In the self-consistent
single-site approximation the coordinate electron Green's function, averaged
over impurity configurations, is given by%
\begin{equation}
G({\boldsymbol{r}}_{1},{\boldsymbol{r}}_{2},\varepsilon )=\sum_{n,k_{y}}\Psi
_{n,k_{y}}^{0\ast }(r_{2})\Psi _{n,k_{y}}^{0}(r_{1})G\left( \varepsilon
,n\right) ,  \label{Gg}
\end{equation}%
where $\Psi _{n,k_{y}}^{0}(r_{1})$ are the 2D electron wave functions in
perpendicular magnetic field,\cite{LL3} and the Green's function $G\left(
\varepsilon ,n\right) $ does not depend on $k_{y}$:%
\begin{equation}
G\left( \varepsilon ,n\right) =\frac{1}{\varepsilon -\hbar \omega _{c}\left(
n+1/2\right) -\Sigma \left( \varepsilon \right) },  \label{Gn}
\end{equation}%
where $\Sigma \left( \varepsilon \right) $ is the electron self-energy part
due to scattering by impurities.

The interlayer conductivity $\sigma _{zz}$, associated with the Hamiltonian (%
\ref{Ht}), can be calculated using the Kubo formula and the formalism,
developed for the metal-insulator-metal junctions.\cite{Mahan} In analogy to
Eq. (44) of Ref. \cite{MosesMcKenzie1999},%
\begin{eqnarray}
\sigma _{zz} &=&\frac{4e^{2}t_{z}^{2}d}{\hbar L_{x}L_{y}}\int d^{2}{%
\boldsymbol{r}}d^{2}{\boldsymbol{r}}^{\prime }\int \frac{d\varepsilon }{2\pi
}\left[ -n_{F}^{\prime }(\varepsilon )\right]  \label{KuboA} \\
&&\times \left\langle \text{Im}G_{R}({\boldsymbol{r}},{\boldsymbol{r}}%
^{\prime },j,\varepsilon )\text{Im}G_{R}({\boldsymbol{r}}^{\prime },{%
\boldsymbol{r}},j+1,\varepsilon )\right\rangle .  \notag
\end{eqnarray}%
The angular brackets in Eq. (\ref{KuboA}) mean averaging over impurity
configurations. Since the impurity distributions on adjacent layers are
uncorrelated, one can perform this averaging separately for each layer. The
averaged Green's functions are translational invariant: $\left\langle G_{R}({%
\boldsymbol{r}},{\boldsymbol{r}}^{\prime },j,\varepsilon )\right\rangle
=\left\langle G_{R}({\boldsymbol{r}}-{\boldsymbol{r}}^{\prime
},j,\varepsilon )\right\rangle $. Therefore, one can perform the integration
over ${\boldsymbol{r}}^{\prime }$, which removes the sample size $L_{x}L_{y}$%
:%
\begin{eqnarray}
\sigma _{zz} &=&\frac{2\sigma _{0}\Gamma _{0}}{\pi \nu _{2D}}\int d^{2}{%
\boldsymbol{r}}\int d\varepsilon \left[ -n_{F}^{\prime }(\varepsilon )\right]
\label{KuboB} \\
&&\times \left\langle \text{Im}G_{R}({\boldsymbol{r}},j,\varepsilon
)\right\rangle \left\langle \text{Im}G_{R}({\boldsymbol{r}},j+1,\varepsilon
)\right\rangle ,  \notag
\end{eqnarray}%
where we introduced the interlayer conductivity without magnetic field%
\begin{equation}
\sigma _{0}=e^{2}t_{z}^{2}\nu _{2D}d/\hbar \Gamma _{0},  \label{s0}
\end{equation}%
$\nu _{2D}=2N_{LL}/\hbar \omega _{c}=m^{\ast }/\pi \hbar ^{2}$ is the 2D DoS
at the Fermi level in the absence of magnetic field per two spin components,
and $N_{LL}$ is the LL degeneracy per unit area.

When the magnetic field is perpendicular to the conducting layers, the
coordinate dependence of the electron Green's function on the adjacent
layers is the same. Then the integration over ${\boldsymbol{r}}$ for the
Green's function of the form (\ref{Gg}) is very simple and gives the factor $%
N_{LL}$:%
\begin{equation}
\sigma _{zz}=\frac{\sigma _{0}\Gamma _{0}\hbar \omega _{c}}{\pi }\int
d\varepsilon \left[ -n_{F}^{\prime }(\varepsilon )\right] \sum_{n}\left\vert
\text{Im}G_{R}(\varepsilon ,n)\right\vert ^{2}.  \label{sp1}
\end{equation}%
In the zero-temperature limit, where $-n_{F}^{\prime }(\varepsilon )=\delta
\left( \varepsilon -\mu \right) $, and in weak magnetic field, where the
summation over $n$ can be replaced by the integration over $n$, Eq. (\ref%
{sp1}) gives%
\begin{equation}
\sigma _{zz}\left( B\right) =\sigma _{0}\Gamma _{0}/\left\vert \text{Im}%
\Sigma \left( \mu ,B\right) \right\vert  \label{sw}
\end{equation}%
in agreement with the standard theory.

\medskip

\textbf{Calculation. }In strong magnetic field, when $\hbar \omega _{c}\gg
\Gamma _{0}=\pi n_{i}U^{2}\rho _{3D}=\pi N_{i}V_{0}^{2}N_{LL}/\hbar \omega
_{c}=\hbar /2\tau _{0}$, one can consider each Landau level separately. In
the self-consistent single-site approximation\cite{Ando} the electron
Green's function on each LL is given by%
\begin{equation}
G\left( E,n\right) =\frac{E+E_{g}\left( 1-c_{i}\right) -\sqrt{\left(
E-E_{1}\right) \left( E-E_{2}\right) }}{2EE_{g}},  \label{GAndo0}
\end{equation}%
and the DoS on each LL is described by the well-known dome-like function\cite%
{Ando}
\begin{equation}
\frac{-\text{Im}G_{R}\left( E,n\right) }{\pi }=D\left( E\right) =\frac{\sqrt{%
\left( E-E_{1}\right) \left( E_{2}-E\right) }}{2\pi \left\vert E\right\vert
E_{g}},  \label{DoSAndo}
\end{equation}%
where the electron energy $E$ is counted from the last occupied LL: $E\equiv
\varepsilon -\varepsilon _{2D}\left( n_{F},k_{y}\right) ,~$and $%
E_{g}=N_{LL}V_{0},$ where the LL degeneracy per unit area is $N_{LL}=1/2\pi
l_{Hz}^{2}=eB/2\pi \hbar c$. The boundaries of the DoS dome in Eq. (\ref%
{DoSAndo}) are%
\begin{equation}
E_{1}=E_{g}\left( \sqrt{c_{i}}-1\right) ^{2},~E_{2}=E_{g}\left( \sqrt{c_{i}}%
+1\right) ^{2},  \label{E12}
\end{equation}%
where $c_{i}$ is the dimensionless ratio of the impurity concentration to
the electron concentration on one LL:
\begin{equation}
c_{i}=N_{i}/N_{LL}=2\pi l_{Hz}^{2}n_{i}d.  \label{ci}
\end{equation}%
The function $D\left( E\right) $ in Eq. (\ref{DoSAndo}) is nonzero in the
interval $0<E_{1}<E<E_{2}$ and normalized to unity: $\int D\left( E\right)
dE=1$. The LL half-width
\begin{equation}
\Gamma _{B}\equiv \left( E_{2}-E_{1}\right) /2=2E_{g}\sqrt{c_{i}}\propto
\sqrt{B}.  \label{GE}
\end{equation}%
The LL broadening $\Gamma _{B}$ in Eq. (\ref{DoSAndo}) is much larger than $%
\Gamma _{0}$ and depends on magnetic field, which is emphasized by the
subscript "$B$". The ratio
\begin{equation}
\Gamma _{B}/\Gamma _{0}\approx \sqrt{4\hbar \omega _{c}/\pi \Gamma _{0}}\gg 1
\label{GEc}
\end{equation}%
grows as $\sqrt{B}$ in high magnetic field.

Taking zero temperature and substituting Eq. (\ref{DoSAndo}) into Eq. (\ref%
{sp1}) we obtain%
\begin{equation}
\sigma _{zz}\left( E\right) =\frac{\sigma _{0}\Gamma _{0}\hbar \omega _{c}}{%
\pi }\sum_{n}\left( \frac{\sqrt{\left( E-E_{1}\right) \left( E_{2}-E\right) }%
}{2\left\vert E\right\vert E_{g}}\right) ^{2},  \label{sSC1}
\end{equation}%
where $E\equiv \mu -\varepsilon _{2D}\left( n_{F},k_{y}\right) $ and the
real part of the square root must be taken, which is nonzero only in the
interval $E_{1}<E<E_{2}$. The monotonic part $\bar{\sigma}_{zz}$ of
conductivity can be obtained by the averaging of Eq. (\ref{sSC1}) over the
oscillation period $\hbar \omega _{c}$:%
\begin{eqnarray}
\bar{\sigma}_{zz} &=&\int_{E_{1}}^{E_{2}}\sigma _{zz}\left( E\right)
dE/\hbar \omega _{c}  \notag \\
&=&\frac{\sigma _{0}\Gamma _{0}}{2\pi E_{g}^{2}}\left[ \frac{E_{2}+E_{1}}{2}%
\ln \left( \frac{E_{2}}{E_{1}}\right) +E_{1}-E_{2}\right]  \notag \\
&=&\frac{2\sigma _{0}\Gamma _{0}}{\pi E_{g}}\left[ \frac{1+c_{i}}{2}\ln
\left( \frac{\sqrt{c_{i}}+1}{\sqrt{c_{i}}-1}\right) -\sqrt{c_{i}}\right] .
\label{sSC}
\end{eqnarray}%
When $c_{i}\gg 1$, this simplifies to
\begin{equation}
\bar{\sigma}_{zz}\approx \frac{2\sigma _{0}\Gamma _{0}}{\pi E_{g}\sqrt{c_{i}}%
}=\sigma _{0}\sqrt{\frac{4\Gamma _{0}}{\pi \hbar \omega _{c}}}.  \label{sSS}
\end{equation}%
The interlayer conductivity in Eq. (\ref{sSS}) decreases with the increase
of magnetic field: $\bar{\sigma}_{zz}\propto B^{-1/2}$. Qualitatively, this
dependence is obtained by substituting $\left\vert \text{Im}\Sigma \left(
\mu ,B\right) \right\vert \approx \Gamma _{B}$ and Eq. (\ref{GEc}) into Eq. (%
\ref{sw}):
\begin{equation}
\bar{\sigma}_{zz}\approx \frac{\sigma _{0}\Gamma _{0}}{\left\vert \text{Im}%
\Sigma \right\vert }\approx \sigma _{0}\frac{\Gamma _{0}}{\Gamma _{B}}%
=\sigma _{0}\sqrt{\frac{\pi \Gamma _{0}}{4\hbar \omega _{c}}}.  \label{sQ}
\end{equation}%
In Ref. \cite{GrigWeaklyIncoherent} the qualitative arguments, similar to
those in the derivation of Eq. (\ref{sQ}), have been applied to show the
monotonic growth and the change in the angular dependence of interlayer
magnetoresistance. However, the arguments in Eq. (\ref{sQ}) are not strict,
because $\left\vert \text{Im}\Sigma \left( \mu ,B\right) \right\vert \neq
\Gamma _{B}$, being a strongly oscillating function of magnetic field $B$
and of Fermi level $\mu $. Therefore, the calculated value of $\bar{\sigma}%
_{zz}$ in Eq. (\ref{sSS}) is $4/\pi \approx 1.27$ times greater than the
qualitative estimate in Eq. (\ref{sQ}), and the above calculation of
interlayer conductivity, resulting in Eq. (\ref{sSS}), is more strict than
in Ref. \cite{GrigWeaklyIncoherent}.

\medskip

\textbf{Discussion.}The physical origin of the decrease of the mean
interlayer conductivity $\bar{\sigma}_{zz}$ can be understood as follows.
The 2D electrons in magnetic field are much stronger affected by the
impurity potential: they become localized, and the energy of each localized
electron state $m$ is shifted by the energy $W\left( m\right) \sim
N_{i}V_{0} $. This energy shift depends on the electron state $m$ and on the
conducting layer $j$. Therefore, when the electron tunnels between two
conducting layers, the energy of the initial and final states are different,
which decreases the interlayer conductivity.

The large increase of the effective imaginary part of the electron self
energy $\left\vert \text{Im}\Sigma \left( \mu ,B\right) \right\vert $ as
compared to $\Gamma _{0}$ in the limit $\hbar \omega _{c}\gg \Gamma
_{0},t_{z}$, resulting to the decrease of the interlayer conductivity
according to Eq. (\ref{sQ}), can also be obtained by the following
qualitative arguments. The average difference $\Delta W\left( m\right) $ of
the energy shifts of two localized electron states is determined by the
fluctuation of the number of impurities effectively interacting with the
localized electron. This number is approximately $N_{i}/N_{LL}=c_{i}>1$, and
the typical fluctuation of this number is $\sim \sqrt{c_{i}}$. The average
difference of the energy shift between two localized states is $\Gamma
_{W}\approx \sqrt{\left\langle \left( \Delta W\right) ^{2}\right\rangle }%
\sim W/\sqrt{c_{i}}\sim \sqrt{\Gamma _{0}\hbar \omega _{c}}$ serves as an
effective $\left\vert \text{Im}\Sigma \right\vert $ in Eq. (\ref{sQ}).
Indeed, the fluctuating shift of the electron energy is equivalent to the
coordinate dependent Re$\Sigma \left( \boldsymbol{r}\right) $ in the
electron Green's function in Eq. (\ref{Gn}). The averaging of the electron
Green's function over impurity configurations is then similar to the
integration over Re$\Sigma \left( \boldsymbol{r}\right) $ with distribution
of the width $\Gamma _{W}$. For the Lorentzian distribution of the energy
shift $W$
\begin{equation}
D\left( W\right) =\Gamma _{W}/\pi \left[ \left( W-\left\langle
W\right\rangle \right) ^{2}+\Gamma _{W}^{2}\right] \label{LorW}
\end{equation}%
this immediately gives the imaginary part $\Gamma _{W}\sim \sqrt{\Gamma
_{0}\hbar \omega _{c}}$ of the electron Green's function:
\begin{eqnarray}
\left\langle G_{R}(\varepsilon ,n)\right\rangle &=&\int \frac{dW\,D\left(
W\right) }{E-W-\varepsilon _{2D}\left( n\right) -i\Gamma _{0}} \notag \\
&=&\frac{1}{E-\left\langle W\right\rangle -\varepsilon _{2D}\left(
n\right) -i\left( \Gamma _{0}+\Gamma _{W}\right) }. \label{GW}
\end{eqnarray}%
In the Green's function in Eq. (\ref{GAndo0}), obtained in the
self-consistent single-site approximation\cite{Ando}, this averaging over
the energy shifts of localized electron states is hidden, and the resulting
value of $\left\vert \text{Im}\Sigma \right\vert \sim $ $\Gamma _{W}\gg
\Gamma _{0}$ in the interval $E_{1}<E<E_{2}$ where the DoS is nonzero.

Eq. (\ref{sSS}) gives the decrease of the monotonic part of
conductivity $\bar{\sigma}_{zz}\propto B^{-1/2}_{z}$. It has a
singularity at $B=0$ because it is derived in the limit of strong
magnetic field when $\hbar \omega _{c}\gg \Gamma _{0}$. In the
crossover region $\hbar \omega _{c}\sim \Gamma _{0}>t_{z}$ the
above arguments remain qualitatively valid, but the quantitative
dependence $\bar{\sigma}_{zz}\left( B\right) $ requires additional
calculation.

In the calculation we assumed the normalized impurity
concentration $c_{i}>1$, because the numerous weak defects and the
impurities, situated far from the conducting layers, are important
for the lifting of LL degeneracy in all layered
materials.\cite{Imp} Therefore, $c_{i}>1$ even in the strongest
pulsed magnets with $B\sim 100T$.

We do not go beyond the self-consistent single-site approximation\cite{Ando}
in studying the influence of the impurity potential, because further
corrections give only the small tails to the DoS distribution.\cite%
{Ando1,Baskin,Brezin,Burmi} Hence, these corrections do not change the main
result. We also disregard the electron-electron interactions, which
restricts our study to the limit when the Fermi energy is much greater than
the cyclotron energy, so that many Landau levels are occupied. The chemical
potential oscillations\cite{ChemPotOscTheory,ChemPotOscTheory1} are also
neglected for two reasons: (i) they do not considerably affect the
nonoscillating part of conductivity and (ii) they are strongly damped
(almost cancelled) by the MQO of the sample volume.\cite{Magnetostriction}
This magnetostriction was directly observed in beryllium.\cite%
{Magnetostriction} No chemical potential oscillations are observed also in
very anisotropic layered organic metals $\beta ^{\prime \prime }$-(BEDT-TTF)$%
_{2}$SF$_{5}$CH$_{2}$CF$_{2}$SO$_{3}$.\cite{W2}

\medskip

To summarize, we calculate the interlayer conductivity in strong magnetic
field in very anisotropic quasi-two-dimensional metals. The calculation is
performed in the framework of the coherent tunnelling model, given by the
Hamiltonian in Eqs. (\ref{H})-(\ref{Vi}). In this calculation the impurity
scattering is considered in the self-consistent single-site approximation,
which is much more accurate for layered almost 2D metals than the
traditionally used Born approximation. This allows to obtain the new
qualitative effect: the strong growth of interlayer magnetoresistance with
the increase of magnetic field along conductivity and perpendicular to the
conducting layers [see Eq. (\ref{sSS})]. This result may explain the
numerous experimental observations in strongly anisotropic layered organic
metals\cite{Zuo1999,Kang,W3,W4,Wang2005,Incoh2009,Wosnitza2002}, where the
interlayer conductivity strongly decreases with the increase of magnetic
field along conductivity in contrast to the standard theory\cite%
{Abrik,Ziman,Yagi1990,MosesMcKenzie1999}.

\medskip

The work was supported by GK P1419 of the FCP program "Nauchnye i
Nauchno-Pedagogicheskie Kadry Rossii" and by the Foundation "Dynasty".

\end{document}